\documentclass[12pt]{article}
\usepackage{amssymb} 
\usepackage{epsfig}
\newcommand{\be}{\begin{equation}}
\newcommand{\ee}{\end{equation}}
\newcommand{\bea}{\begin{eqnarray}}
\newcommand{\eea}{\end{eqnarray}}

\begin{document}

\begin{center}

{\bf The possible  test of the calculations of
nuclear matrix elements of the $(\beta \beta)_{0\nu}$-decay}

\end{center}

\begin{center}
S. M. Bilenky \footnote{ The present address: INFN, Sez. di Torino and Dip. di Fisica Teorica,
Univ. di Torino, I-10125 Torino, Italy}

\vspace{0.3cm}
{\em IFAE, Facultat de Ciencies, Universitat Autonoma
de Barcelona, 08193, Bellaterra, Barcelona, Spain \\}

\vspace{0.3cm} {\em  Joint Institute
for Nuclear Research, Dubna, R-141980, Russia\\}
\end{center}
\begin{center}
J. A. Grifols\\
\vspace{0.3cm}
{\em IFAE, Facultat de Ciencies, Universitat Autonoma
de Barcelona, 08193, Bellaterra, Barcelona, Spain \\}

\end{center}

\begin{abstract}
The existing calculations of the nuclear matrix elements
of the neutrinoless double $\beta$-decay
differ by about a factor three. This uncertainty prevents
quantitative interpretation of the results of experiments
searching for this process.
We suggest here that the observation of the neutrinoless double $\beta$-decay
of {\em several} nuclei 
could allow to test calculations of the nuclear matrix elements 
through the comparison of the ratios of the calculated lifetimes
with experimental data.
It is shown that the ratio of the lifetimes   
is very sensitive to different models.

\end{abstract}

\section{Introduction}

The compelling evidences in favour of neutrino oscillations were obtained in the
Super-Kamiokande \cite{S-K}, SNO \cite{SNO} 
and other atmospheric and solar  
neutrino experiments. These findings 
 mean that neutrino masses are different from 
zero 
and fields of the flavour neutrinos are mixture of the left-handed components of the fields of neutrinos with definite masses. 
It is a general consensus that small neutrino masses and neutrino mixing
is a first evidence for new physics.

There are many unsolved problems in the physics of massive and mixed 
neutrinos. The most fundamental one
is the problem of the {\em nature} of neutrinos with definite masses: are they Dirac or
Majorana
particles? The answer to this question can not be obtained via the investigation of neutrino oscillations.
In order to probe the nature of the massive neutrinos it is
necessary to study processes in which the total lepton number is not conserved.
The most sensitive to the possible violation of the total lepton number process is neutrinoless double $\beta$ -decay ($(\beta\beta)_{0\nu}$ -decay) of even-even nuclei.

The data of many experiments on the search for $(\beta\beta)_{0\nu}$ -decay
are available at present (see \cite{Cremonesi,Zdesenko}).        
No any indications in favour of $(\beta\beta)_{0\nu}$ -decay were obtained up to
now.\footnote{ The recent
claim \cite{Klap01} of some evidence of the $ (\beta \beta)_{0\,\nu}$-decay,
obtained from the reanalysis of the data of the Heidelberg-Moscow experiment,
was strongly criticised in  \cite{FSViss02,bb0nu02}.}

The strongest limits on the lifetime of the $(\beta\beta)_{0\nu}$ -decay
were obtained in the Heidelberg-Moscow \cite{HM} and IGEX \cite{IGEX}
 $^{76} \rm{Ge}$ experiments:

\be
T^{0\nu}_{1/2}(^{76} \rm{Ge})\geq 1.9 \cdot 10^{25}\, y\,~~ 
(\rm{H-M});\,~T^{0\nu}_{1/2}(^{76} \rm{Ge})\geq 1.57 \cdot 10^{25}\, y\,~~ 
(\rm{IGEX})\,.
\label{001}
\ee

There are several mechanisms of the neutrinoless double $\beta$-decay.
We will consider here $(\beta\beta)_{0\nu}$ -decay in the framework of the
Majorana neutrino mixing
\be
\nu_{lL} = \sum_{i} U_{li} \nu_{iL},
\label{002}
\ee
where $U$ is Pontecorvo-Maki-Nakagava-Sakata unitary mixing matrix and $\nu_{i}$ is the field of the Majorana neutrino with mass $m_{i}$.
After recent evidences for neutrino oscillations this mechanism appears as the most natural one. 
\footnote{ Other mechanisms of the $(\beta \beta)_{0\nu}$ -decay 
are based on SUSY R-parity 
violating models \cite{Mohap}, on a model with admixture of heavy neutrinos to the light ones
\cite{Fedor} etc. In \cite{Fedor} possibilities to distinguish different 
mechanisms are considered. The proposed tests require
detection of the $(\beta \beta)_{0\nu}$ -transition into excited states
and precise calculations of the transition probabilities.}

In the case of the Majorana neutrino mixing
the matrix element of the
$(\beta\beta)_{0\nu}$ -decay is proportional to the effective Majorana mass
(see \cite{Doi,BPet})
\be
<m> = \sum_{i} U^{2}_{ei}\,m_{i}.
\label{003}
\ee

From the results of $^{76} \rm{Ge}$ experiments
it was found
\be
|<m>|\leq (0.35-1.24)\,~\rm{eV}\,~~(\rm{Heidelberg-Moscow})\,,
\label{004}
\ee

\be
|<m>|\leq (0.33-1.35)\,~\rm{eV}\,~~\rm{IGEX}\,.
\label{005}
\ee

In (\ref{004}) and (\ref{005}) different calculations of nuclear matrix elements were used.

Many new experiments on the search for $(\beta\beta)_{0\nu}$ -decay
of different nuclei
are under preparation in different laboratories.
In these experiments much higher sensitivities
to the effective Majorana mass $|<m>|$ than the present-day ones are 
expected (see \cite{Cremonesi}).
For example, the sensitivities to $|<m>|$ which are planned to be reached in the experiments CUORE ($^{130} \rm{Te}$) \cite{CUORE}, GENIUS
 ($^{76} \rm{Ge}$) \cite{GENIUS} , MAJORANA ($^{76} \rm{Ge}$)
\cite{MAJORANA},
EXO  ($^{136} \rm{Xe}$) \cite{EXO}, MOON ($^{100} \rm{Mo}$) \cite{MOON} 
are, respectively, equal to
$ 2.7\cdot10^{-2}\,\rm{eV}$, $1.5\cdot10^{-2}\,\rm{eV}$,
$2.5\cdot10^{-2}\,\rm{eV}$, $5.2 \cdot10^{-2}\,\rm{eV}$, 
$3.6\cdot10^{-2}\,\rm{eV}$.\footnote{In the calculation of these
sensitivities the nuclear matrix elements, given in \cite{Staudt}, were used.}

The observation of the $(\beta\beta)_{0\nu}$ -decay
would be a proof that neutrinos with definite masses are Majorana particles.
It was shown in many papers (see \cite{BGGKP}
and references therein) that {\em the 
measurement of
the effective Majorana mass $|<m>|$} would allow to obtain an unique information about neutrino mass spectrum and Majorana CP phase.

There exist, however, a serious problem of the determination of
$|<m>|$ from experimental data. It is connected with nuclear matrix elements:
the calculated matrix elements vary within factor
three.

In this note we would like to propose a possible test  
of the calculations of the nuclear matrix elements, based on
the comparison of the results of the calculations with the experimental data.
In order to realize the proposed test it is necessary to observe
$(\beta\beta)_{0\nu}$ -decay of {\em several nuclei}.

\section{Possible test of nuclear matrix elements calculations}

In the framework of 
the Majorana neutrino mixing (\ref{002})
the total probability of the
 $ (\beta \beta)_{0\,\nu}$ - decay has the following general form
(see, \cite{Doi,BPet}):
\be
\Gamma^{0\,\nu}(A,Z) = |<m>|^{2}\,|M(A,Z)|^{2}\,G^{0\,\nu}(E_{0},Z)\,,
\label{006}
\ee
where$ M(A,Z)$ is the nuclear matrix element and
$G^{0\,\nu}(E_{0},Z)$
is  known phase- space factor ($E_{0}$ is the energy release).
Thus, in order to determine  $|<m>|$ from the experimental data we need
to know the nuclear matrix element $M(A,Z)$. This last quantity 
must be calculated.

There exist at present large uncertainties in the calculations of the
nuclear matrix elements of the $ (\beta \beta)_{0\,\nu}$-decay 
(see \cite{Faessler,Suhonen,Elliott}).
Two basic approaches to the calculation are  
used: 
quasiparticle random phase approximation and the nuclear shell model.
Different calculations of the
lifetime of the $ (\beta \beta)_{0\,\nu}$-decay differ
by about one order of magnitude.
For example, for the lifetime of the $(\beta \beta)_{0\,\nu}$ - decay of $^{76}\rm{Ge}$
it was obtained the range  \cite{Elliott} \footnote{The values given in 
(\ref{007})
were calculated under the assumption that $|<m>|= 5\cdot 10^{-2}\,\rm{eV}$}:

\be
6.8\cdot 10^{26}\leq T^{0\,\nu}_{1/2}(^{76}\rm{Ge})\leq 70.8\cdot 10^{26}\,\rm{years}
\label{007}
\ee

The problem of the calculation of the nuclear matrix elements
of the neutrinoless double $\beta$-decay is a real theoretical challenge.
It is obvious that without 
solution of this problem
the effective Majorana neutrino mass $|<m>|$
can not
be determined from the experimental data with reliable accuracy
(see discussion in \cite{Barger}).

We will propose here
a method which allows to check the results
of the calculations
of the nuclear matrix elements of the $ (\beta \beta)_{0\,\nu}$-decay
of different nuclei by confronting them with experimental data.
 We will take into account
the following

\begin{enumerate}

\item 
For small neutrino masses
($m_{i}\lesssim 10\,\rm{MeV}$) the nuclear matrix elements do not
depend on neutrino masses \cite{Doi,BPet}.

\item 
The sensitivity
$|<m>| \simeq \rm{a\, few } 10^{-2}\, eV$ is planned to be reached in
 experiments on the search for neutrinoless double $\beta$ -decay
of {\em different} nuclei.

\end{enumerate}

From (\ref{006}) we have 

\be
R(A,Z/A',Z') = \frac{T^{0\,\nu}_{1/2}(A,Z)}{T^{0\,\nu}_{1/2}(A',Z')}=
\frac{|M(A',Z')|^{2}\,G^{0\,\nu}(E'_{0},Z')}{|M(A,Z)|^{2}\,G^{0\,\nu}(E_{0},Z)}
\label{008}
\ee

Thus, if  the neutrinoless double $\beta$ -decay of  different nuclei
will be observed, the calculated ratios of the corresponding
nuclear matrix elements- squared can be confronted with the experimental values.

In the Table I we present 
the ratios of lifetimes of the $(\beta\beta)_{0\,\nu}$-decay
of several nuclei, calculated in six
different models.
For the lifetimes we used the  values given in \cite{Elliott}. As it is seen from Table I, the calculated ratios
are very sensitive to the model: they vary
within about one order of magnitude.

\begin{center}
 Table I
\end{center}
\begin{center}
The results of the calculation of the ratios of the lifetime of
$(\beta\beta)_{0\,\nu}$-decay of several nuclei in six different models.
 The references
to the corresponding papers are given in brackets.
\end{center}

\begin{center}
\begin{tabular}{|ccccccc|}
\hline
Lifetime ratios
&
\cite{HS84}
&
\cite{Caurier99}
&
\cite{EVZ88}
&
\cite{Staudt}
&
\cite{TS95}
&
\cite{Pantis96}
\\
\hline
$R(^{76}\rm{Ge}/^{130}\rm{Te})$
&
11.3
&
3
&
20
&
4.6
&
3.6
&
4.2
\\
$R(^{76}\rm{Ge}/^{136}\rm{Xe})$
&

&
1.5
&
4.2
&
1.1
&
0.6
&
2
\\
$R(^{76}\rm{Ge}/^{100}\rm{Mo})$
&

&

&
14
&
1.8
&
10.7
&
0.9
\\
\hline
\end{tabular}
\end{center}

As we can see from the Table I, 
the ratio $R(^{76}\rm{Ge}/^{130}\rm{Te})$, calculated in \cite{Staudt} and
\cite{TS95} is equal, correspondingly,
4.6 and 3.6. It is clear that it will be difficult to distinguish 
models \cite{Staudt} and  \cite{TS95} by the observation of the
neutrinoless double $\beta$ -decay of  
$^{76}\rm{Ge}$ and $^{130}\rm{Te}$.
 However, 
it will be no problems to distinguish the corresponding models via the
observation of the  
$(\beta\beta)_{0\,\nu}$-decay of $^{76}\rm{Ge}$ and $^{100}\rm{Mo}$
(the corresponding ratio is equal 1.8 and 10.7, respectively).
This example  illustrates the importance of the investigation of 
$(\beta\beta)_{0\,\nu}$-decay
of more than two nuclei.
The nuclear part of the matrix element of the $(\beta\beta)_{0\,\nu}$-decay
is determined by the matrix element 
of the T-product of two hadronic charged currents 
connected
by the propagator of massless boson (see, for example, \cite{BPet}) 
\footnote{ The relation 
$$ \sum_{i}U^{2}_{ei}\frac{1+\gamma_{5}}{2}\,\frac{\gamma q + m_{i}}{q^{2}-m_{i}^{2}}\,
\frac{1-\gamma_{5}}{2}\simeq <m>\, \frac{1}{q^{2}}\,\frac{1-\gamma_{5}}{2}$$
is used}

\be
\int <A, Z+2\,|T(J^{\alpha}(x_{1})\, J^{\beta}(x_{2})\,|A,Z>\,\frac{e^{-iq(x_{1}-x_{2})}}{q^{2}}\,e^{ip_{1}x_{1}+ip_{2}x_{2}}\,d^{4}x_{1}\,d^{4}x_{2}\,
d^{4}q 
\label{009}
\ee
This matrix element can not be connected with matrix element of any observable
hadronic process. We believe that the method, proposed here, which is based on the 
factorisation of neutrino and nuclear parts of the matrix element of the  
$(\beta\beta)_{0\,\nu}$-decay, is the only possibility to test the 
calculations of the nuclear matrix elements
in a model independent way.

We would like to finish with the following remark. If the ratio
(\ref{008}), calculated in some model, is in agreement with
experimental data this could only mean that the model is correct
up to a possible factor, which does not depend on A and Z (and 
drops out from the ratio (\ref{008})). Such factor
was found and calculated in Ref.\cite{SPVF}.
In this paper in addition to the usual axial and vector terms
in the nucleon matrix element pseudoscalar and weak magnetic formfactors were taken into account.
It was shown that in the case of the light Majorana neutrinos these 
additional terms lead to
a universal $\simeq$ 30 \% reduction of the nuclear matrix elements 
of the $ (\beta \beta)_{0\,\nu}$-decay, which practically does not depend on 
type of the nuclei. This reduction will cause the corresponding raise
of the value of the effective Majorana mass $|<m>|$ that could be obtained 
from the results of the future experiments.

\section{Conclusion}
The observation of the neutrinoless double $\beta$ -decay
would have a great impact on the understanding of the origin of neutrino masses and mixing. The accurate measurement of the effective Majorana mass $|<m>|$
would allow to make important conclusions on the neutrino mass spectrum and
Majorana CP phase (see \cite{BGGKP} and references therein).

Let us consider the minimal scheme of three-neutrino mixing
and label neutrino masses in such a way that $m_1 < m_2< m_3 $.\footnote{ The
LSND result \cite{LSND}, which requires more than three massive and mixed neutrinos,
needs confirmation. The MiniBooNE experiment \cite{MiniB}, started recently,
aims to check
the LSND claim.}
From the results of the neutrino oscillation experiments
only neutrino mass-squared differences $\Delta m^{2}_{21} =m^{2}_{2}-m^{2}_{1}$
and $\Delta m^{2}_{32} =m^{2}_{3}-m^{2}_{2}$
can be inferred. In order to illustrate the importance of the 
measurement of $|<m>|$ 
we will consider three typical neutrino mass spectra,
compatible with the results of neutrino oscillation experiments.

\begin{enumerate}

\item
The hierarchy of neutrino masses $m_1 \ll m_2 \ll m_3 $.

For the effective Majorana mass we have in this case the bound

\be
|<m>|\leq \sin^{2}\,\theta_{\rm{sol}}\,\sqrt{\Delta m^{2}_{\rm{sol}}}+ |U_{e3}|^{2}\,\sqrt{\Delta m^{2}_{\rm{atm}}}\,.
\label{010}
\ee

Using the best-fit values of the neutrino oscillation parameters
in the most favourable MSW LMA region \cite{SNO}
$
\Delta m^{2}_{\rm{sol}}= 5.0 \cdot 10^{-5}
\rm{eV}^{2};\,~~\tan^{2}\theta_{\rm{sol}}= 0.34\,
$, the value of the atmospheric neutrino mass-squared difference 
$
\Delta m^{2}_{\rm{atm}}= 2.5 \cdot 10^{-3}
\rm{eV}^{2}$,
obtained from the analysis of the data of the Super-Kamiokande
atmospheric neutrino experiment 
\cite{S-K}, and the CHOOZ \cite{CHOOZ} bound 
\be
|U_{e3}|^{2}\leq 4\cdot 10^{-2}
\label{011}
\ee
for the effective Majorana mass we have 
\be
|<m>| \leq 3.8 \cdot 10^{-3}\,\rm{eV}\,,
\label{012}
\ee
This bound is significantly smaller than the expected sensitivity
of the future $(\beta\beta)_{0\nu}$ -experiments.\footnote {Let us note, however, that at the next stage of the GENIUS experiment 
(10 tons of enriched $^{76}\rm{Ge}$) the bound (\ref{012})
is expected  to be reached \cite{Klopp}}. 
 
Thus, the observation of the $(\beta\beta)_{0\nu}$- decay in the 
experiments of the next generation
would presumably create a problem for the hierarchy of neutrino mass,
motivated by the famous see-saw mechanism of neutrino mass generation.

\item
Inverted hierarchy of neutrino masses: $m_{1}\ll m_{2}<m_{3}$.

The effective Majorana mass is given in this case by 

\be
|<m>|\simeq  \left( 1 - \sin^{2}2\,\theta_{\rm{sol}}\,\sin^{2}\alpha \right)
^{\frac{1}{2}}\,\sqrt{\Delta m^{2}_{\rm{atm}}}\,,
\label{013}
\ee

where $\alpha =\alpha_3 -\alpha_2 $ is the difference of the Majorana CP phases. Using the best- fit value of the parameter 
$\tan^{2}\theta_{\rm{sol}}$ we have

\be
\frac{1}{2}\,\sqrt{\Delta m^{2}_{\rm{atm}}}
\lesssim|<m>|
\lesssim\sqrt{\Delta m^{2}_{\rm{atm}}}\,,
\label{014}
\ee

Thus, in the case of the inverted mass hierarchy
the scale of $|<m>|$ is determined
by $\sqrt{\Delta m^{2}_{\rm{atm}}}\backsimeq 5\cdot 10^{-2}\rm{eV}$.
If the value of $|<m>|$ is in the range (\ref{013}),
which can be reached in the future experiments 
on the search for  
$(\beta \beta)_{0\nu}$ -decay,
it will be a signature of inverted neutrino mass hierarchy.

\item

Practically degenerate neutrino mass spectrum.

If $m_1 \gg \sqrt{\Delta m^{2}_{\rm{atm}}}$ for neutrino masses we have
$m_2 \simeq m_3\simeq m_1$. Effective Majorana mass in this case is equal to
\be
|<m>| \simeq m_{1}\,~|\sum_{i=1}^{3}U_{ei}^{2}|.
\label{015}
\ee
Taking into account the CHOOZ bound (\ref{011}), in the case of 
the LMA solution of the solar neutrino problem
we have $|U_{e3}|^{2}\ll |U_{e1}|^{2}, |U_{e2}|^{2}$. 
Hence, we can neglect the contribution of $|U_{e3}|^{2}$ to the effective
Majorana mass. From (\ref{015}) we have
\be
m_{1}\simeq \frac{|<m>|}{\left( 1 - \sin^{2}2\,\theta_{\rm{sol}}\,\sin^{2}\alpha \right)
^{\frac{1}{2}}}\,.
\label{016}
\ee
For the best-fit LMA value $\tan^{2}\theta_{\rm{sol}}=0.34$ 
from (\ref{016}) we obtain the bounds

\be
|<m>|\leq m_{1}\lesssim 2\,|<m>|
\label{017}
\ee

Thus, if it will occur that the effective Majorana mass $|<m>|$ is 
significantly larger than $\sqrt{\Delta m^{2}_{\rm{atm}}}$
it will be an evidence for the practically degenerate neutrino mass spectrum.

\end{enumerate}

The measurement of the effective Majorana mass  $|<m>|$
could allow to obtain an information about the  Majorana CP phase difference
$\alpha$. In fact 
we have \cite{BGKP}.

\be
\sin^{2}\alpha \simeq \left( 1 -\frac{ |<m>|^{2}}{m^{2}_{0}}
\right)\,~\frac{1}{\sin^{2}2\,\theta_{\rm{sol}}}\,,
\label{018}
\ee
where $m_{0}= \sqrt{\Delta m^{2}_{\rm{atm}}}$ in the case of the inverted 
neutrino mass hierarchy and  $m_{0}=  m_{1}$
in the case of practically degenerate mass spectrum.
In the case of the CP conservation in the lepton sector
$\sin^{2}\alpha =0$ ( $\sin^{2}\alpha =1$) for equal (opposite) CP parities of $\nu_{2}$ and $\nu_{3}$.

Thus, accurate measurement of $|<m>|$ ($\Delta m^{2}_{\rm{atm}}$ and
$\sin^{2}2\,\theta_{\rm{sol}}$ )
would allow to determine Majorana CP phase difference $\alpha$ in the case
of the inverted hierarchy of neutrino masses.

In order to determine the parameter $\sin^{2}\alpha $
in the case of the degenerate neutrino mass spectrum 
we need to know 
$m_1$.
The mass $m_1$ can be inferred from experiments on the measurement of the high-
energy part of the $\beta$-spectra.
From the latest data of Mainz \cite{Mainz} and Troitsk \cite{Troitsk} tritium
experiments the bound 
$
m_1 \leq 2.2 \, \rm{eV}
$ was obtained.
In the future tritium experiment KATRIN \cite{Katrin} the sensitivity
$
m_1 \simeq 0.35\, \rm{eV}
$
is expected.

An information about absolute values of neutrino masses can be obtained
also from cosmological data. From 2dF Galaxy Redshift survey
and CMB data it was found that \cite{Elgaroy}
$
\sum_{i}m_{i} \simeq (1.8-2) \, \rm{eV}
$.
The future MAP/PLANK CMB data and high precision 
Sloan Digital Sky Survey
could render
\cite{Hu}
$
\sum_{i}m_{i} \simeq 0.3 \, \rm{eV}\,.
$

In conclusion we would like to stress that in order to obtain
an unique information on neutrino mass spectrum and Majorana CP phase
from the observation 
of the neutrinoless double $\beta$- decay  
we need to have a possibility to control the calculations of the nuclear matrix elements.
We have shown here that 
the observation of the  $(\beta\beta)_{0\,\nu}$-decay of
{\em several} nuclei would allow to test in a model independent way 
the results of calculations of the nuclear matrix elements.

S.M.B. acknowledges the support of the ``Programa de Profesores Visitantes de IBERDROLA de Ciencia y Tecnologia''.

\end{document}